\documentclass[twocolumn,aps,prl,superscriptaddress,floatfix,longbibliography]{revtex4-1}  

\usepackage[draft]{changes}
\usepackage[normalem]{ulem}

\usepackage{graphicx}
\usepackage{amssymb} 
\usepackage{dcolumn}
\usepackage{bm}

\usepackage[mathlines]{lineno}
\usepackage[colorlinks,linkcolor=blue,anchorcolor=blue,citecolor=blue,urlcolor=blue]{hyperref}
\usepackage{multirow}
\usepackage{color}
\usepackage{amsmath}

\makeatletter
\renewcommand*{\@fnsymbol}[1]{\ensuremath{\ifcase#1\or \dagger\or *\or \ddagger\or
\mathsection\or \mathparagraph\or \|\or **\or \dagger\dagger \or
\ddagger\ddagger \else\@ctrerr\fi}} \makeatother
\usepackage{color}

\usepackage[draft]{changes}
\usepackage[normalem]{ulem}

\usepackage{color}%

\begin{document}
%
\title{Bonding Hierarchy and Coordination Interaction Leading to High Thermoelectricity in Wide Bandgap TlAgI$_2$}


\author{Xiaoying Wang}
\affiliation{State Key Laboratory for Mechanical Behavior of Materials, School of Materials Science and Engineering,
             Xi'an Jiaotong University, Xi'an 710049, China} 

\author{Mengyang Li}
\affiliation{School of Physics, Xidian University, Xi'an710071, China}

\author{Minxuan Feng}
\affiliation{State Key Laboratory for Mechanical Behavior of Materials, School of Materials Science and Engineering,
             Xi'an Jiaotong University, Xi'an 710049, China}

\author{Xuejie Li}
\affiliation{State Key Laboratory for Mechanical Behavior of Materials, School of Materials Science and Engineering,
             Xi'an Jiaotong University, Xi'an 710049, China}

\author{Yuzhou Hao}
\affiliation{State Key Laboratory for Mechanical Behavior of Materials, School of Materials Science and Engineering,
             Xi'an Jiaotong University, Xi'an 710049, China}


\author{Wen Shi}
\affiliation{School of Chemistry, Sun Yat-sen University, Guangzhou, Guangdong, 510006, China}
\affiliation{Institute of Green Chemistry and Molecular Engineering, Sun Yat-sen University, Guangzhou, Guangdong, 510006, China}

\author{Jiangang He}
\affiliation{School of Mathematics and Physics, University of Science and Technology Beijing, Beijing 100083, China}



                
\author{Xiangdong Ding}
\affiliation{State Key Laboratory for Mechanical Behavior of Materials, School of Materials Science and Engineering,
             Xi'an Jiaotong University, Xi'an 710049, China}

\author{Zhibin Gao}
\email[E-mail: ]{zhibin.gao@xjtu.edu.cn}
\affiliation{State Key Laboratory for Mechanical Behavior of Materials, School of Materials Science and Engineering,
             Xi'an Jiaotong University, Xi'an 710049, China}




\date{\today}
\begin{abstract} 
High thermoelectric properties are associated with the phonon-glass electron-crystal paradigm. Conventional wisdom suggests that the optimal bandgap of semiconductor to achieve the largest power factor should be between 6 and 10$\kappa_BT$. To address challenges related to the bipolar effect and temperature limitations, we present findings on Zintl-type TlAgI$_2$, which demonstrates an exceptionally low lattice thermal conductivity of 0.3 W m$^{-1}$ K$^{-1}$ at 300 K. The achieved figure of merit ($ZT$) for TlAgI$_2$, 
featuring a 1.55 eV bandgap, reaches a value of 2.20 for p-type semiconductor.
This remarkable $ZT$ is attributed to the existence of extended antibonding states [Ag-I] in the valence band. Furthermore, the bonding hierarchy, influencing phonon anharmonicity, and coordination bonds, facilitating electron transfer between the ligand and the central metal ion, significantly contribute to electronic transport. 
This finding serves as a promising avenue for the development of high $ZT$ materials with wide bandgaps at elevated temperatures.

\end{abstract}

\maketitle


\section{I. Introduction}

Thermoelectric technology, offering clean and sustainable means, can directly and reversibly convert heat into electrical energy. Typically, the thermoelectric conversion efficiency is gauged by the dimensionless figure of merit, $ZT = {S^2 \sigma T}/{\kappa}$, where $S$, $\sigma$, $\kappa$, and $T$ represent the Seebeck coefficient, electrical conductivity, thermal conductivity, and working temperature, respectively. However, these parameters are tightly interconnected, and improving $ZT$ necessitates optimizing the adversely interdependent $S$, $\sigma$, and $\kappa$ as a collective. Therefore, there are several degrees of freedom to enhance $ZT$, such as spin, orbital, charge, and lattice~\cite{he2017advances,snyder2008complex}.


In a given working temperature range, the optimal $ZT$ is constrained by the intrinsic electronic bandgap. Many celebrated narrow bandgap thermoelectric materials, such as PbTe ($E_g$=0.28 eV)~\cite{poudeu2006highAngew,albanesi2000electronic} and (Bi,Sb)$_2$Te$_3$ ($E_g$=0.13 eV)~\cite{poudel2008highSci}, have been identified. However, the thermoelectric properties are significantly affected when there is a substantial number of both electrons and holes contributing to charge transport, known as bipolar charge transport. This phenomenon occurs when electrons are excited across the bandgap, producing minority charge carriers (e.g., holes in an n-type material) in addition to majority charge carriers (e.g., electrons in an n-type material). Bipolar effects are observed in small bandgap materials at high temperatures ($k_B$$\sim$$E_g$). Consequently, the Seebeck coefficient is dramatically affected because the minority charge carriers add a Seebeck voltage of the opposite sign to the majority carriers, greatly reducing the thermopower $\left| S \right|$~\cite{li2020positive,chen2021leveraging,germanese2022bipolar,pei2012thermoelectric,hoang2010impurity,C1JM13888J}. Moreover, narrow-gap semiconductors cannot be effectively utilized at higher temperatures.

By employing a good rule of thumb, $E_g = 2 e S_{max} T $~\cite{xiao2020seeking}, where $S_{max}$ is the maximum Seebeck coefficient, and $e$ is the unit charge, wide bandgap semiconductors could mitigate the bipolar effect and temperature range limitations~\cite{2014Ultralow,zhao2016ultrahigh,liu2009improved}. In other words, wide bandgap semiconductors have the potential to overcome the restrictions of narrow bandgap materials and serve as promising thermoelectric candidates. For example, the three-element Heusler Li$_2$NaSb, with a 1.72 eV bandgap, achieves a $ZT$ value of 1.20~\cite{xing2017electronic}. A copper-tin compound, Cu$_2$Se, with a 1.20 eV bandgap, exhibits a $ZT$ value of 1.40~\cite{lu2015multiformity}. The stannide tin compounds Cu$_2$ZnSnSe$_4$, featuring a 1.44 eV bandgap, demonstrate a $ZT$ value of 0.75~\cite{shi2009thermoelectric,Liu2009AWP}. However, all these systems have $ZT$ values below 2.0, primarily due to poor electrical properties and high lattice thermal conductivity ($\kappa_L$).

In this study, we leverage chemical bonding hierarchy and coordination interaction to enhance the transport properties of the wide bandgap material TlAgI$_2$. The concept of chemical bond hierarchy involves ionic bonding, covalent bonding, and coordination interaction~\cite{acharyya2023extended, pal2018bonding,wan2022bonding}, which explains the coexistence of weak and rigid bonds within materials. In materials undergoing thermally induced large amplitude vibrations, such as La$_2$Zr$_2$O$_7$~\cite{2020Vibrational} and Bi$_4$O$_4$SeCl$_2$~\cite{tong2023glass}, the intrinsic coexistence of rigid crystalline sublattices and fluctuating noncrystalline sublattices is observed. This atomic-level heterogeneity results in vibrational modes that generate a mismatch in the phonon density of states, thereby enhancing phonon anharmonicity and reducing $\kappa_L$~\cite{qiu2014part}.


We discovered that Zintl-type TlAgI$_2$ exhibits an ultralow $\kappa_L$ of 0.30 W m$^{-1}$ K$^{-1}$ at 300 K, achieved by considering quartic anharmonicity renormalization and the off-diagonal term of the heat flux operators. Additionally, the weakening of bonds and strong phonon-phonon interactions are attributed to the antibonding states just below the Fermi level in the electronic band structure, arising from interactions between silver $4d$ and iodine $5p$ orbitals. Moreover, the unexpectedly strong hole transport performance, characterized by a large hole density of states, is influenced by the coordination interactions forming a stable coordination complex, Ag-I~\cite{skoug2011role}. The wide bandgap, coupled with high energy band asymmetry, counteracts bipolar effects, resulting in a notably high Seebeck coefficient up to 704 $\mu$V K$^{-1}$ at a hole concentration of 10$^{18}$ at 1200 K. Ultimately, the achieved $ZT$ values for TlAgI$_2$ with a 1.55 eV bandgap reach 2.20 and 1.82 for p-type and n-type concentrations, respectively. This finding suggests the potential of using bonding hierarchy and coordination interactions in designing high-temperature thermoelectric materials with wide bandgaps.

\section{II. COMPUTATIONAL METHODS}
Generally, $\kappa_L$ is the summation of the Peierls contribution from diagonal term $\kappa_p$ and the Wigner distribution
from off-diagonal term $\kappa_c$~\cite{simoncelli2019unified}, $\kappa_{Total}$ = $\kappa_c$ + $\kappa_p$. Wherein the $\kappa_c$ originates from Wigner distribution associated with the wave-like tunnelling~\cite{kane2012zener,simoncelli2019unified} and loss of coherence between different vibrational eigenstates. The $\kappa_c$ can be expressed as,
\begin{eqnarray}
\begin{split}
\label{eqn1}
\kappa_c  =  & \frac{\hbar^2}{k_B T^2 V N_0} \sum_{\bm{q}} \sum_{s \neq s'} \frac{ \omega^s_{\bm{q}} + \omega^{s'}_{\bm{q}} }{2}  \bm{v_q^{s, s'}  v_q^{s', s}}\\
& \times \frac{ \omega^s_{\bm{q}}  n^s_{\bm{q}} ( n^s_{\bm{q}} + 1 ) + \omega^{s'}_{\bm{q}} n^{s'}_{\bm{q}} ( n^{s'}_{\bm{q}}  +1 )  }{ 4 ( \omega^{s'}_{\bm{q}} - \omega^s_{\bm{q}}  )^2  + ( \Gamma^s_{\bm{q}} + \Gamma^{s'}_{\bm{q}} )^2 }\\
& \times ( \Gamma^s_{\bm{q}} + \Gamma^{s'}_{\bm{q}} ),
\end{split}
\end{eqnarray}
where $\hbar$, $k_B$, $T$, $V$, $N_0$ are the reduced Planck constant, %
Boltzmann constant, absolute temperature, primitive cell volume, and 
the total number of sampled phonon wave vectors in the first Brillouin 
zone, respectively.
For the Peierls-Boltzmann transport equation, the diagonal contribution $\kappa_p$ can be calculated as, %
\begin{eqnarray}
\label{eqn2}
\kappa_p=\frac{\hbar^2}{k_B T^2 V N_0} \sum_{\lambda} n_\lambda (n_\lambda + 1) \omega^2_\lambda \bm{v}_\lambda \otimes \bm{v}_\lambda \bm{\tau}_\lambda,
\end{eqnarray}
where $\kappa_p$ represents the $\kappa_p^{3,4ph}$ considering anharmonic phonon renormalization (APRN) at  finite temperatures, three-phonon (3ph) and four-phonon (4ph) scatterings. %
It is derived from Peierls contribution related to the particle-like propagation of phonon wave packets. %
$n_\lambda$, $\omega_\lambda$,  $v_\lambda$, and $\tau_\lambda$
are the equilibrium component
of the phonon population, frequency, group velocity, and lifetime for the 
$\lambda$ mode (wave vector \textbf{$q$} and branch index $s$), respectively. %
Except for $\tau_\lambda$, all the above parameters can be obtained from harmonic 
approximation (HA). We adapted 3ph and 4ph scattering from the 
self-consistent phonon (SCPH)~\cite{werthamer1970self} theory to obtained 
$\tau_\lambda$ including anharmonic effects beyond perturbation 
theory that considers the quantum effect of phonons~\cite{debernardi1995anharmonic, gao2018unusually,wang2023role}. %

Among various existing approaches such as self-consistent ab initio lattice dynamics (SCAILD)~\cite{souvatzis2008entropy} 
and stochastic self-consistent harmonic approximation (SSCHA)~\cite{errea2014anharmonic}, self-consistent phonon (SCPH) approximation is one effective method that can rigorously account for the first-order correction of phonon 
frequencies from the quartic anharmonicity. The SCPH approach can better describe the 
soft phonon modes and strong anharmonicity. In brief, the SCPH can be 
written as~\cite{tadano2015self,xia2020high}
\begin{eqnarray}
\label{eqn3}	 
{\Omega}_{\lambda}^{2} = {\omega}_{\lambda}^2+2{\Omega}_{\lambda}\sum\limits_{{\lambda}_{1}} I_{\lambda\lambda_1},
\end{eqnarray}
where $\omega_\lambda$ is the original phonon frequency from 
the harmonic approximation and $\Omega_\lambda$ is the 
temperature-dependent renormalized phonon frequency. The scalar 
$I_{\lambda\lambda_1}$ can be obtained by,
\begin{eqnarray}
\label{eqn4}	 
{I_{\lambda\lambda_1}}=\frac{\hbar}{8 N_0} \frac{V^{(4)} (\lambda,-\lambda,\lambda_1,-\lambda_1)}{\Omega_{\lambda}\Omega_{\lambda_1}} \left[1+2n_\lambda(\Omega_{\lambda_1})\right],
\end{eqnarray}
in which $V^{(4)}$ is the fourth-order IFCs in the reciprocal representation and 
phonon population $n_\lambda$ satisfies Bose-Einstein distribution as a function 
of temperature. Both Eq. (\ref{eqn3}) and Eq. (\ref{eqn4}) have parameters 
$I_{\lambda\lambda_1}$ and $\Omega_\lambda$ in common, and thus SCPH equation can 
be solved iteratively. Note that $I_{\lambda\lambda_1}$ can be interpreted as the 
interactions between a pair of phonon modes, $\lambda$, and $\lambda_1$ including 
the temperature effects~\cite{tadano2015self, xia2020high}. %

DFT calculations were performed using the \textit{Vienna ab initio simulation 
package} (VASP)~\cite{kresse1996efficient} with the projector-augmented 
wave (PAW) method~\cite{blochl1994projector, kresse1999ultrasoft}. We used 
the PBEsol functional to obtain lattice constants. Cutoff energy 
of 400 eV was used with 11 $\times$ 11$\times$ 11 Monkhorst-Pack 
$\textit{k}$-grids. The self-consistent iteration for the energy convergence criterion 
was  10$^{-8}$~eV, and all geometries were optimized by the conjugate-gradient 
method until none of the residual Hellmann-Feynman forces exceeded 10$^{-6}$ eV/\AA. %
The optimized conventional cell lattice constant of tetragonal 
I4/mcm phase (No. 140), a=b=8.188 \AA, c=7.562 \AA. A 2 $\times$ 2 $\times$ 2 supercell and 5 $\times$ 5 $\times$ 5 $\textit{k}$-points were employed in all finite displacement 
calculations. %

We generated force-displacement data by performing $\textit{ab initio}$ molecular dynamics (AIMD) simulation with a 2 $\times$ 2 $\times$ 2  supercell at 300 K for 2000 steps with a time step of 2 fs using a Nosé-Hoover thermostat and $10^{-6}$ eV energy threshold. We sampled 40 atomic configurations that were equally spaced in time by removing the first 400 steps from the trajectories and then randomly displaced all of the atoms within the supercell by 0.02 \AA~(second-order) and 0.1 \AA~(higher-order) in random directions in each configuration to decrease cross-correlations in the sensing matrix formed by products of atomic displacements. Finally, the 40 uncorrelated sets were computed using accurate DFT calculations with a $10^{-8}$ eV energy convergence threshold. 
We used 12 $\times$ 12 $\times$ 12 ngrids for $\kappa_p^{3ph}$, and 7 $\times$ 7 $\times$ 7 for $\kappa_p^{3,4ph}$ and $\kappa_c$. 
The electronic band structure and crystal orbital Hamilton population (COHP) were calculated using a 9 $\times$ 9 $\times$ 9 $\textit{k}$-meshes. %
The elastic constants,  dielectric constants, deformation potential, and wave functions 
were gained with 12 $\times$ 12 $\times$ 12 $\textit{k}$-meshes.
The carrier transport properties were
obtained in uniform 41 $\times$ 41 $\times$ 45 $\textit{k}$-point grids in electronic transport~\cite{ganose2021efficient}. %

We systematically studied the effect of quartic anharmonicity on the lattice dynamics, electronic transport, and thermal transport properties of TlAgI$_2$ by leveraging recent advances
including (i) compressive sensing lattice dynamics (CSLD)~\cite{zhou2014lattice, zhou2019compressive, zhou2019compressive2} to establish the high-order inter-atomic force constants (IFCs), that utilized the compressive sensing technique~\cite{4472240} to select the physically relevant IFCs from the force-displacement data under the constraints enforced by the space group symmetry. (ii) rigorous calculations of temperature-dependent phonons used SCPH theory and higher-order multiphonon scattering rates~\cite{tadano2015self, feng2016quantum}, (iii) evaluation of  $\kappa_L$ employed a unified theory that accounts simultaneously for diagonal term from the standard Peierls contribution and off-diagonal terms from the coherent Wigner distribution~\cite{simoncelli2019unified, isaeva2019modeling}. 
%
(iv) The Seebeck coefficient, conductivity, and power factor were calculated by considering the electron-phonon coupling such as the acoustic deformation potential, ionized impurity, and polar optical phonon scattering, as implemented in the amset package~\cite{ganose2021efficient}.

\section{III. RESULTS AND DISCUSSION}

\begin{figure}
\includegraphics[width=1.0\columnwidth]{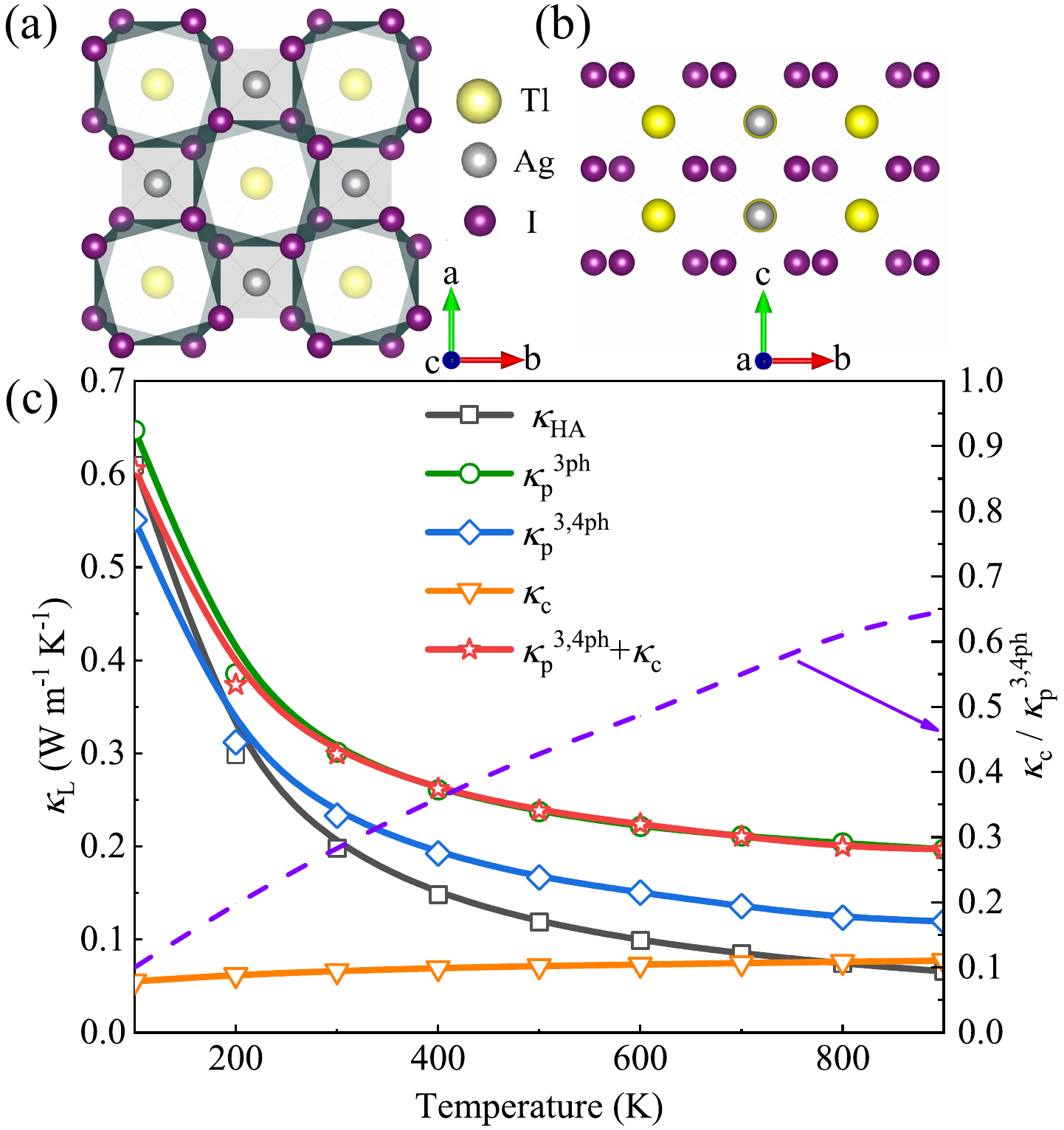}
\caption{The crystal structure of the conventional cell of TlAgI$_2$ is depicted, containing 16 atoms. 
The top view (a) and side view (b) along the crystallographic $c$-axis and $a$-axis are shown.
(c) ``HA'' denotes the harmonic approximation. 
$\kappa_p^{3,4ph}$ represents lattice thermal conductivity, considering quartic anharmonic phonon renormalization (APRN) and four-phonon (4ph) interactions. $\kappa_c$ is the contribution of the off-diagonal terms from the Wigner distribution. Thallium (Tl), silver (Ag), and iodine (I) atoms are represented by yellow, gray, and purple colors, respectively. The right vertical axis indicates the ratio of $\kappa_c$/$\kappa_p^{3,4ph}$. The lattice thermal conductivity ($\kappa_L$) is averaged over the three coordinates.
%
\label{fig1}}
\end{figure}

TlAgI$_2$ adopts a tetragonal structure (space group I4/mcm [140]), where Tl, Ag, and I occupy the 4a, 4b, and 8h sites with a total of 8 atoms in the primitive cell. In this structure, Tl and I form an octahedral cage~\cite{jana2017intrinsic}, illustrated in Fig.~\ref{fig1}(a), with the Ag element embedded within the cage. All outcomes take into account the Self-Consistent Phonon (SCPH) effect, except for the Harmonic Approximation (HA). The lattice thermal conductivity ($\kappa_L$) is averaged over the three crystalline directions. $\kappa_p^{3,4ph}$ represents lattice thermal conductivity considering quartic anharmonic phonon renormalization (APRN) and four-phonon (4ph) interactions. 

The influence of SCPH is crucial, as evidenced by the contrast between HA and $\kappa_p^{3ph}$, highlighting a pronounced temperature effect on phonons. Compared with the $\kappa_p^{3ph}$ value of 0.30 W m$^{-1} K^{-1}$, $\kappa_p^{3,4ph}$ decreases to 0.23 W m$^{-1} K^{-1}$ at 300 K due to additional 4ph scattering. However, when considering the contribution of the off-diagonal term ($\kappa_c$), the total lattice thermal conductivity increases to 0.30 W m$^{-1} K^{-1}$, constituting $\kappa_p^{3,4ph}$ + $\kappa_c$. Interestingly, the contribution of coherent phonons $\kappa_c$ grows significantly with increasing temperature. At room temperature, the lattice thermal conductivity of $\kappa_p^{3ph}$ aligns with that of $\kappa_p^{3,4ph}$ + $\kappa_c$. 

\begin{figure*}
%
\includegraphics[width=2.0\columnwidth]{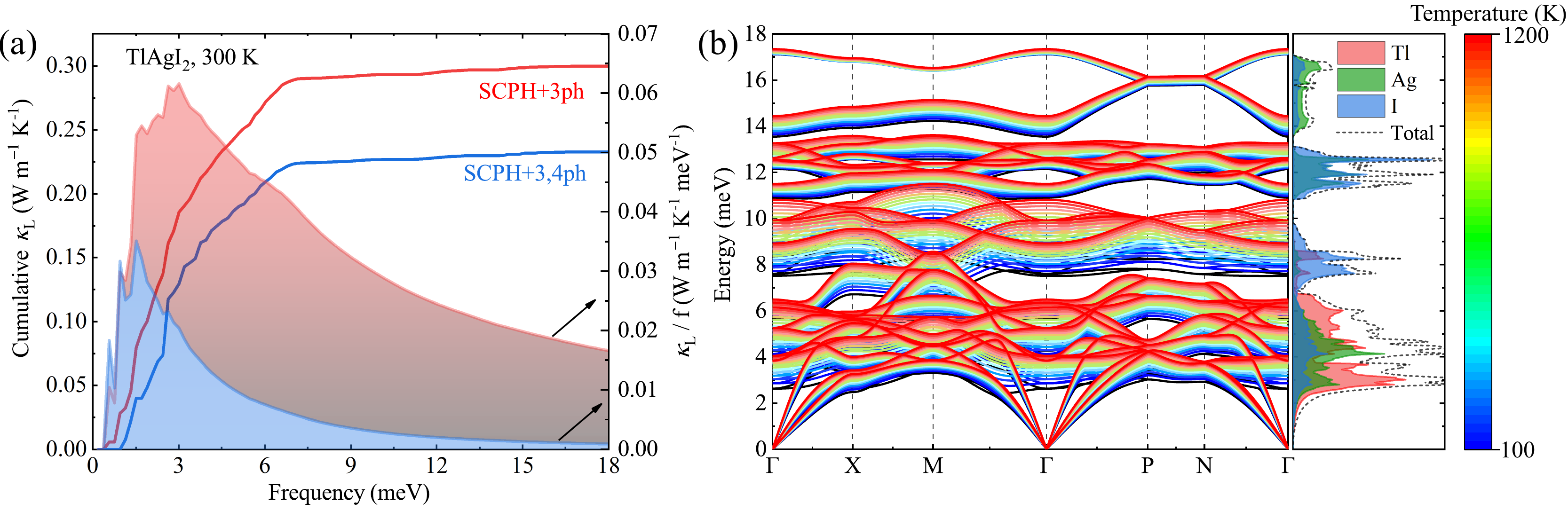}
\caption{ (a) The frequency-resolved $\kappa_L$ represented by the filled area below the lines, and the cumulative $\kappa_L$ spectrum, indicated by the line. The methods are SCPH+3ph (red) and SCPH+3,4ph (blue) for TlAgI$_2$ material at 300 K, respectively. (b) The temperature-dependent colourful phonon spectrum for TlAgI$_2$ from T = 100 K to 1200 K using SCPH. The black line at T = 0 K denotes the harmonic approximation (HA).  The element-decomposed density of states (PDOS) are shown in the red, green, and blue colors representing thallium (Tl), silver (Ag), and iodine (I) atoms. Total density of state is illustrated in dotted line. 
%
\label{fig2}}
\end{figure*}

Subsequently, we delve into the frequency-resolved analysis (filled region) and cumulative trends (solid lines) of $\kappa_L$ at 300 K to further scrutinize the microscopic mechanisms of phonon vibrations leading to low $\kappa_L$, as illustrated in Fig.~\ref{fig2}(a). The $\kappa_L$ spectrum and the cumulative $\kappa_L$ with respect to frequency reveal that the primary contributors to $\kappa_L$ are the phonon branches within the 2-4 meV range, affirming the validity of our 4ph scattering calculation. At the same time, we find that acoustic phonons below 4 meV are the main contribution to the lattice thermal conductivity, no matter the $\kappa_p^{3ph}$ (red line) and $\kappa_p^{3,4ph}$ (blue line).



The temperature-dependent phonon dispersion unquestionably reveals the stiffening of both acoustic and optical branches with increasing temperature probably originating with weakly coupling between high-frequency optical phonons and overdamped acoustic phonons (I$_{\lambda\lambda_1}$ is positive as mentioned in Eq. (\ref{eqn4})), as depicted in Fig.~\ref{fig2}(b). The vibrational spectra of Tl atoms predominantly occupy the 2-4 meV regime, exerting significant influence on thermal transport as evidenced by the atom-projected phonon density of states (PDOS). It is also affirmed that the vibrations of Tl atoms serve as the primary scattering channel. Moreover, the strongly interlinked phonon branches within the low-frequency range of 4-5 meV are anticipated to establish a substantial phonon-phonon scattering channel for both acoustic and low-frequency optical modes~\cite{lin2021ultralow}.


Finally, and perhaps most crucially, the outermost layer of Tl comprises three electrons, including 6s$^2$ and 6p$^1$. Theoretically, the valence states can be monovalent, divalent, and trivalent. Conversely, Ag possesses only one electron in the outermost layer, 5p$^1$. This electron likely transfers to I, forming a stable ionic bond. Consequently, Tl can contribute only one valence electron to I, leaving a 6s$^2$ lone pair electron. The Bader charge, detailed in Table SII in the Supplemental Materials, 
further supports this. As a result, the compound exhibits the following valence state: Tl$^{1+}$Ag$^{1+}$(I$^{1-}$)$_2$. 
%

The monovalent Tl in the TlAgI$_2$ system~\cite{jana2017intrinsic} with a 6s$^2$ lone pair, exhibits overlapping wave functions of the lone electron pair with nearby valence electrons. This overlap causes a nonlinear repulsive electrostatic force as atoms approach each other, leading to off-centering of the atoms. The interaction of the lone electron pair originating from the Tl 6s orbital with adjacent atoms induces bond anharmonicity and significantly reduces $\kappa_L$~\cite{skoug2011role,jana2017intrinsic,mukhopadhyay2020ultralow, rathore2019orig, morelli2008intrinsically}.


\begin{figure*}
\includegraphics[width=2.0\columnwidth]{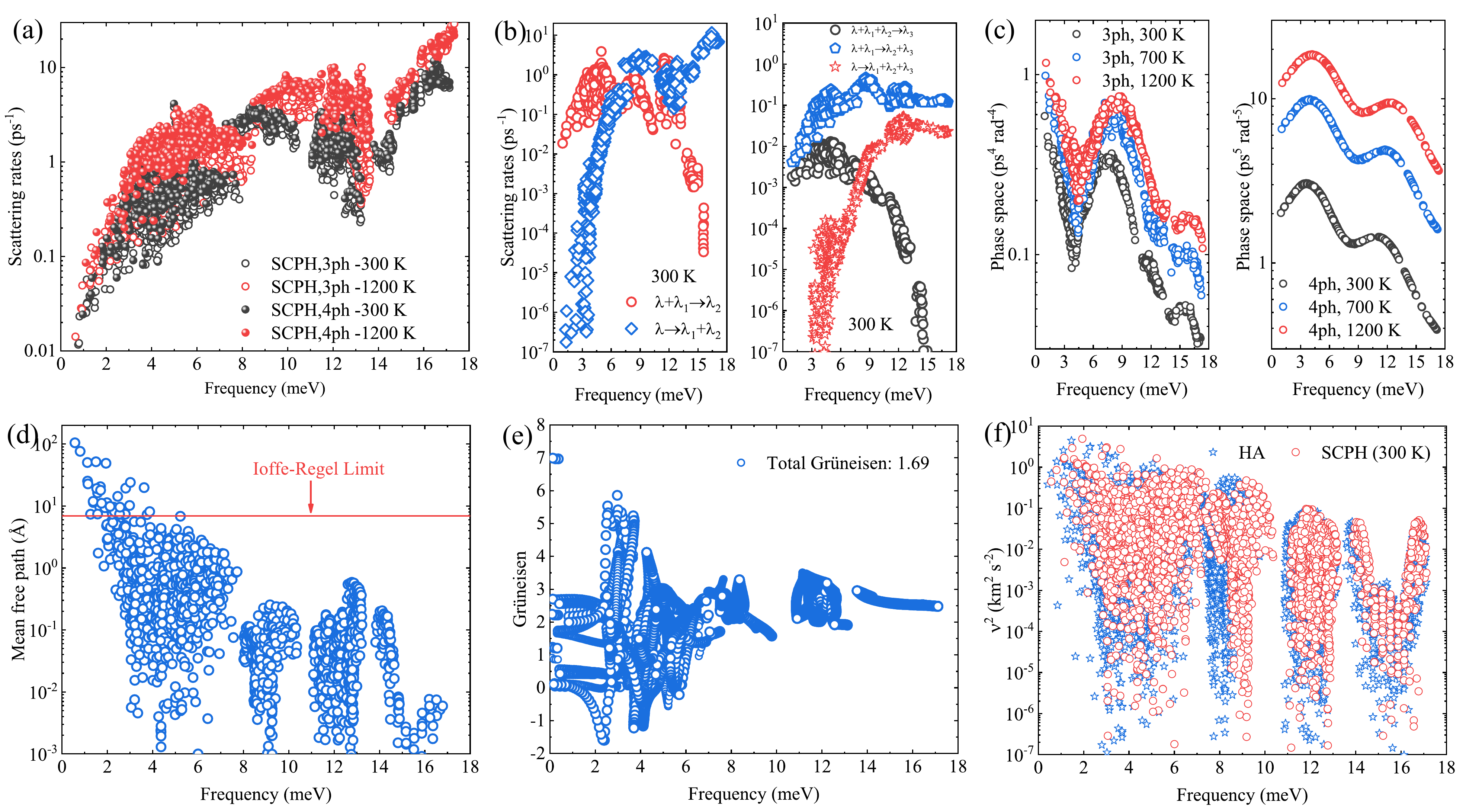}
\caption{
%
Thermal transport mechanism analysis of ultralow $\kappa_L$ of TlAgI$_2$. (a) Phonon-phonon scattering rates of 3ph and 4ph at both 300 K and 1200 K. (b) Description of the decomposed scattering of 3ph and 4ph into splitting, reconstruction, and combination processes. (c) Illustration of the phonon scattering phase space of 3ph and 4ph. (d) The mean free path of phonons with the Ioffe-Regel Limit~\cite{hussey2004universality} at 300 K. (e) The Grüneisen parameter at 300 K. (f) The square of the phonon group velocity $v^2$ in the harmonic approximation and anharmonic phonon renormalization at 300 K.
\label{fig3}}
\end{figure*}

\begin{figure*}
\includegraphics[width=2.0\columnwidth]{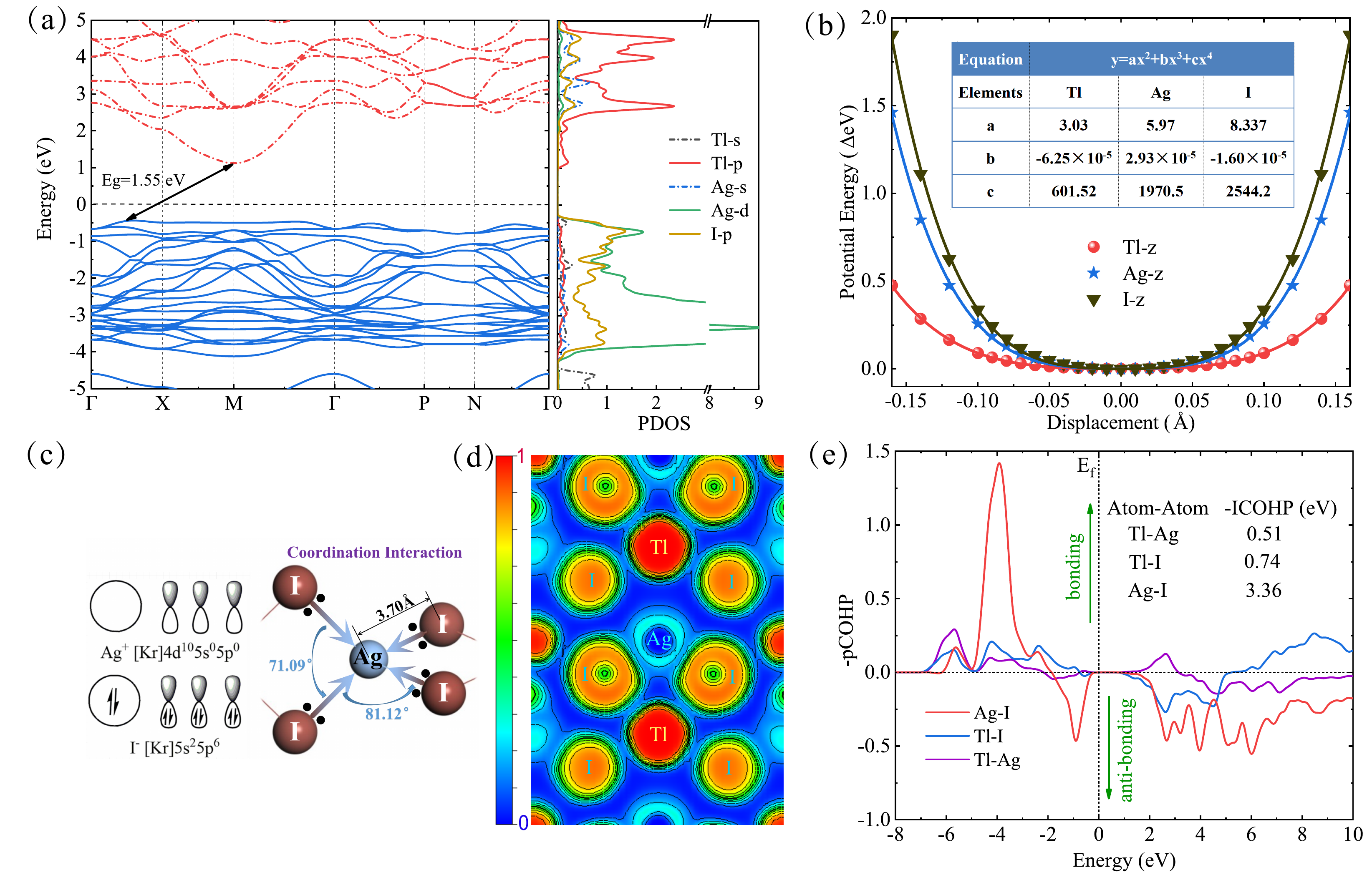}
\caption{Electrical transport mechanism analysis of TlAgI$_2$. (a) The electron energy band and projective density of states for different elemental orbitals in TlAgI$_2$. (b) The potential energy versus displacement along crystallographic $z$-direction of TlAgI$_2$ in conventional cell. (c) The coordination states of silver and iodine, with brown and blue atom representing Ag and I, respectively. Bond angles are 81.12$^\circ$ and 71.09$^\circ$ (The symmetric position possesses same bond Angle), all the bond lengths are 3.70\AA. (d) The 2D electron location function (ELF) of TlAgI$_2$. (e) The calculated projected crystal orbital Hamilton population (pCOHP) analysis for the Tl-Ag, Tl-I, and Ag-I of TlAgI$_2$, where the negative values of -pCOHP represent bonding states, and the positive values indicate anti-bonding states. The energy is shifted to the Fermi level at 0 eV. 
\label{fig4}}
\end{figure*}


All available three-phonon (3ph) and four-phonon (4ph) scattering rates are depicted in Fig.~\ref{fig3}(a). At the same temperature, the relationship of scattering rates (SR) can be expressed as SR$_{4ph}$ $\ge$ SR$_{3ph}$, indicating that 4ph scattering is not negligible in this system. Furthermore, scattering rates increase with temperature. Therefore, by including 4ph scattering, the lattice thermal conductivity is generally smaller than when considering only 3ph scattering, resulting in lower $\kappa_L$ with increasing temperature.

Fig.~\ref{fig3}(b) displays the absorption and emission processes of the 3ph and 4ph as a function of frequency at 300 K. For the 3ph scattering, we consider the phonon 
splitting ($\lambda {\to} \lambda_1 + \lambda_2$) and combination ($\lambda 
+ \lambda_1 {\to}  \lambda_2$). In the case of 4ph, we count for phonon splitting ($\lambda {\to} \lambda_1 + \lambda_2 + \lambda_3$), combination ($\lambda + \lambda_1 + \lambda_2 {\to}  \lambda_3$), as well as
redistribution ($\lambda + \lambda_1 {\to} \lambda_2 + \lambda_3$) processes. %
In the low-frequency region dominated by acoustic modes, 3ph combination is stronger than the splitting situation, while the redistribution process of 4ph is dominant. However, in the high-frequency region dominated by optical modes, the splitting process of 3ph becomes more important.  
For 4ph scattering, the splitting course dominates the scattering process and has the same order as the redistribution process. A similar phenomenon was observed in the case of the halide perovskite material CsPbBr$_3$~\cite{wang2023role}.


The phonon phase space at different temperatures for 3ph and 4ph interactions reveals a strong temperature dependence, particularly for 4ph, as illustrated in Fig.~\ref{fig3}(c). The phase space of both 3ph and 4ph scattering increases as the temperature rising from 300 K to 700 K and 1200 K. It is important to note that the units of phase space for 3ph and 4ph are different, preventing a direct comparison between them~\cite{2017Four}.


We observed that crystalline materials with an antibonding state and bonding hierarchy exhibit a coexistence of population phonons and coherent phonons. Coherent phonons are likely to manifest in the presence of complex crystals~\cite{simoncelli2019unified}, which require a large unit cell with numerous closely spaced phonon branches, strong anharmonicity (where phonon linewidths exceed interbranch spacings)~\cite{xia2020microscopic}, and phonons below the Ioffe-Regel Limit (mean free path is around the interatomic spacing) contributing to heat transport due to their wavelike ability to interfere and tunnel~\cite{simoncelli2022wigner}, as depicted in Fig.~\ref{fig3}(d). Given that most mean free paths are smaller than the minimum atomic distance, known as the Ioffe-Regel Limit, it suggests that $\kappa_c$ is likely important in TlAgI$_2$~\cite{simoncelli2022wigner}.

%

To deepen our understanding of the physical mechanism behind the ultralow $\kappa_L$ and the significance of anharmonicity, we calculate the Grüneisen parameter, as illustrated in Fig.~\ref{fig3}(e). The extent of anharmonicity is typically measured by the Grüneisen parameters ($\gamma$). In the top panel of Fig.~\ref{fig3}(e), relatively large values of $\gamma$ are observed in intertwined portions of the acoustic and optical branches regime at 300 K, confirming stronger scattering within the frequency range of 2 to 4 meV. 
This is closely linked to the heavy atomic mass of the Tl element and the bonding hierarchy. 
Moreover, TlAgI$_2$ exhibits substantial anharmonicity with a total Grüneisen value of 1.69, indicating more anharmonicity than the PbTe value of 1.45 ~\cite{slack1979thermal}. Fig.~\ref{fig3}(f) demonstrates the temperature effect on $v^2$, where $v$ represents the group velocity. We observe a low speed of sound at 300 K for TlAgI$_2$, stemming from the heavy atomic mass of the Tl element and the vibration resistance, as indicated by the renormalized phonon dispersions at finite temperature.


\begin{figure*}
\includegraphics[width=2.0\columnwidth]{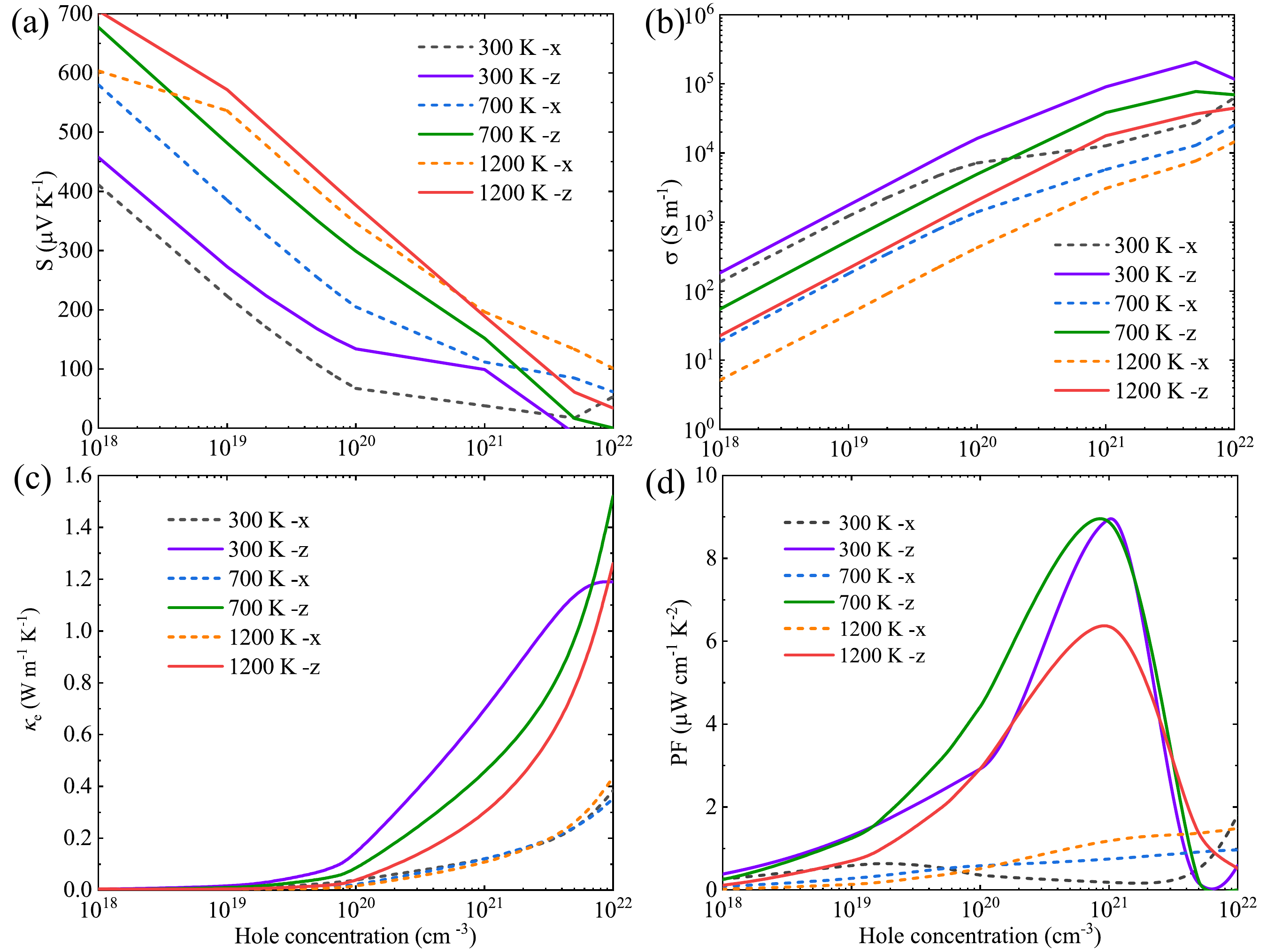}
\caption{%
%
The carrier transport performance of TlAgI$_2$. (a) The Seebeck coefficient, (b) Electrical conductivity, (c) Electronic thermal conductivity, and (d) Power factor $PF$ ($PF=S^2 \sigma$) of the p-type with $x$-axis and $z$-axis, respectively.
\label{fig5}}
\end{figure*}


In Fig.~\ref{fig4}(a), the electronic band structure and projective density of states for different elemental orbitals of TlAgI$_2$ are presented. TlAgI$_2$ is identified as an indirect bandgap semiconductor, where the conduction band minimum (CBM) is located at the M point, and the valence band maximum (VBM) is situated at the $\Gamma$ to X high symmetry line. The computed bandgap using the PBEsol functional is 1.55 eV, consistent with the findings from the Material Project database mp-27801. 
TlAgI$_2$ is characterized as a wide bandgap semiconductor, a common feature in high temperature thermoelectric semiconductor materials, such as half-heusler alloys FeNb$_{0.88}$Hf$_{0.12}$Sb~\cite{fu2015realizing}.
Specifically, the CBM is predominantly contributed by the Tl atom's 6p orbital, while the VBM is almost entirely attributed to the Ag 4d and I 5p orbitals. 


The conduction band near the Fermi energy level exhibits pronounced valleys, while the valence band appears notably flat. In semiconductors with highly asymmetric bands near the Fermi energy, there is a substantial difference in the density of states between electrons and holes. When the concentration of one type of carrier significantly exceeds that of the other, the detrimental impact of the bipolar effect on the Seebeck coefficient can be effectively mitigated~\cite{gong2016investigation}. Simultaneously, a higher valence density of states enhances the material's responsiveness to the effects of thermoelectric conversion.


For an in-depth understanding of atomic-level dynamics, we computed potential energy curves by displacing atoms along the $z$-direction relative to their static equilibrium positions at the $\Gamma$ point. As depicted in Fig.~\ref{fig4}(b), Ag and I atoms are confined in steep potential wells, while the Tl atom resides in a flat potential well. This suggests that the Tl atom can readily oscillate within the surrounding hollow cage with a large amplitude. The electrostatic repulsion between the localized electrons of Tl and the neighboring atoms likely induces significant vibrations of Tl. Consequently, chemical bonds various strengths coexist in the system, the pronounced mismatch in their vibrational modes impedes phonon propagation.
This increased anharmonicity effectively reduces the $\kappa_L$~\cite{yue2023strong}. The flat potential well of Tl is consistent with its large Atomic Displacement Parameter (ADP) . The loose bonding between Tl and other elements, coupled with electrostatic repulsion between the localized electrons (lone pair 6$s^2$) of Tl and the surroundings, plausibly drives substantial displacement of Tl, leading to a large ADP along the $c$-axis, as illustrated in the Supplemental Materials FIG. S2.


The schematic of coordination compounds Ag$^{1+}$I$^{1-}$, illustrated in Fig.~\ref{fig4}(c), reveals the unique electron configurations of the components. The I$^{1-}$ anion adheres to the eight-electron rule, possessing four lone pair electrons. Conversely, the Ag$^{1+}$ cation displays unpaired 5s and 5p orbitals. This distinctive electron arrangement leads to the formation of the coordination bond I$^{1-}$-Ag$^{1+}$, aligning with the Projected Density of States (PDOS) where p orbitals of Ag and I significantly contribute to the valence band. The coordination interactions are strengthened by similar energy levels and matching symmetry of the orbitals involved in bond formation. This results in the establishment of a stable coordination complex, AgI, where the I$^{1-}$ anion acts as a ligand, and the Ag$^{1+}$ cation serves as the central metal ion. Such coordination bonds play a crucial role in electrical transport processes, facilitating the transfer of electrons between the ligand and the central metal ion~\cite{skoug2011role}. 


\begin{figure*}
\includegraphics[width=2.0\columnwidth]{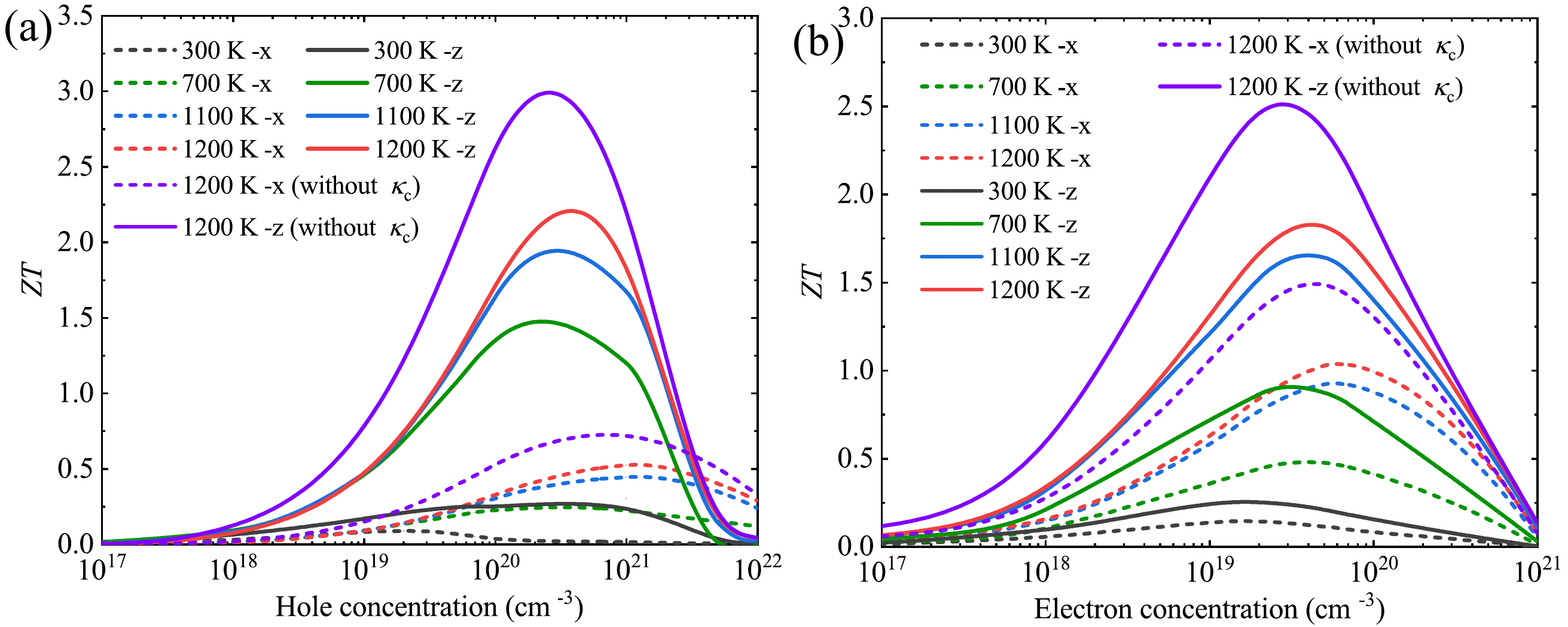}
\caption{%
%
The calculated thermoelectric figure of merit ($ZT$) with the lattice thermal conductivity ($\kappa_L$) originating from both phonon ($\kappa_p$) and coherent ($\kappa_c$) contributions ($\kappa_L =\kappa_p^{3,4ph} + \kappa_c$) for (a) p-type and (b) n-type TlAgI$_2$ at temperatures of 300 K, 700 K, 1100 K, and 1200 K. A comparison is provided by calculating $\kappa_p$ only for TlAgI$_2$ at 1200 K for both p-type and n-type carriers.
\label{fig6}}
\end{figure*}

Furthermore, the exceedingly weak bonding between Tl and I, particularly evident around the Tl element and the delocalization of Tl, as depicted in Fig.~\ref{fig4}(d), suggests a spherical charge density due to the electron lone pair of Tl (refer to Fig.~\ref{fig2}(b)). This implies that the interactions between Tl and the surrounding Ag and I atoms are primarily electrostatic. With a combination of ionic bonding, covalent bonding, and coordination interaction, the hierarchical bonding and mismatched bond energies result in strong anharmonic interactions, especially below the Fermi level. Due to the strong electron separability at the top of the valence band, the robust coordination interactions of Ag and I contribute to a high density of states effective mass~\cite{acharyya2023extended, pal2018bonding}. 


Fig.~\ref{fig4}(e) illustrates the projected crystal orbital Hamilton population (pCOHP) analysis. The Ag-I interaction exhibits strong bonding states, with antibonding states below the Fermi level in the electronic structure resulting from the interactions between silver 4d and iodine 5p orbitals. This weakens the bond and induces strong phonon anharmonicity. In contrast, Tl-Ag and Tl-I interactions display weak bonding states. Evidently, Tl atoms are loosely bound to other atoms, attributed to the lone pair and rattling vibration mode. The value of -ICOHP represents the integrals from -infinity to the Fermi level of COHP, signifying the ability to form bonds between different atoms, consistent with the 2D electron location function shown in Fig.~\ref{fig4}(d).


The hole transport performance for p-type TlAgI$_2$ is depicted in Fig.~\ref{fig5}. Observations reveal that the Seebeck coefficient $S$ decreases with increasing carrier concentration $n$ of holes at the same temperature, while $S$ increases with increasing $T$ at the same $n$. The electrical conductivity $\sigma$ of p-type TlAgI$_2$ shows a positive correlation with the carrier concentration $n$ and a negative correlation with the temperature $T$. Notably, p-type TlAgI$_2$ exhibits a larger $S$ due to high energy band asymmetry and a wide bandgap that counteracts the bipolar effect, especially at high temperatures, as illustrated in Fig.~\ref{fig4}(a). The conductivity $\sigma$ is influenced by the carrier concentration $n$ and inversely proportional to the temperature $T$. The former is attributed to the increasing concentration $n$, contributing to the conductivity, while the latter is due to the boosting of the electron-phonon interaction scattering rate with increasing temperature.


In general, the electrical conductivity $\sigma$ and electronic thermal conductivity $\kappa_e$ exhibit a similar trend with the increase in hole concentration, as depicted in Fig.~\ref{fig5}(b) and Fig.~\ref{fig5}(c), consistent with the Wiedemann-Franz law ($\kappa_e$ = ${L \sigma T}$)~\cite{jonson1980mott}. Due to the coexistence of relatively large Seebeck coefficient $S$ and $\sigma$, a high thermoelectric power factor ($PF$) is achieved, for instance, 8.94 $\mu$W cm$^{-1}$ K$^{-2}$ at 300 K in the $z$-direction at a hole concentration ($n_h$) of $10^{21}$ cm$^{-3}$. Notably, there is a positive correlation between the bandgap and the temperature at the highest $ZT$ value~\cite{xiao2020seeking}. Considering the bandgap of TlAgI$_2$ is 1.55 eV, which is relatively large, we tentatively selected the highest temperature to be 1200 K, commonly used for high-temperature thermoelectric materials. It is observed that the highest power factor occurs at a hole doping concentration of $10^{21}$. The electrical transport performance for n-type TlAgI$_2$ is presented in the Supplemental Material and shows lower performance.


As depicted in Fig.~\ref{fig6}, the $ZT$ values for p-type doping are relatively high, reaching 2.20, whereas for n-type doping, it is only 1.82. The highest $ZT$ for n-type doping is observed at a carrier concentration of 2$\times$10$^{19}$ at 1200 K, and the most reasonable concentration for p-type doping is 2$\times$10$^{20}$ at 1200 K. An interesting observation is that the $ZT$ without considering $\kappa_c$ can be enhanced from 2.20 to 3.00, even though the value of $\kappa_c$ is only 0.08 W m$^{-1}$ K$^{-1}$ (as shown in Supplemental Material TABLE. SI.) at 1200 K. Therefore, it is crucial to consider $\kappa_c$ when calculating the $ZT$ to avoid overestimating the thermoelectric performance, especially for materials with low thermal conductivity~\cite{tong2023glass}.

\begin{figure}
\includegraphics[width=1.0\columnwidth]{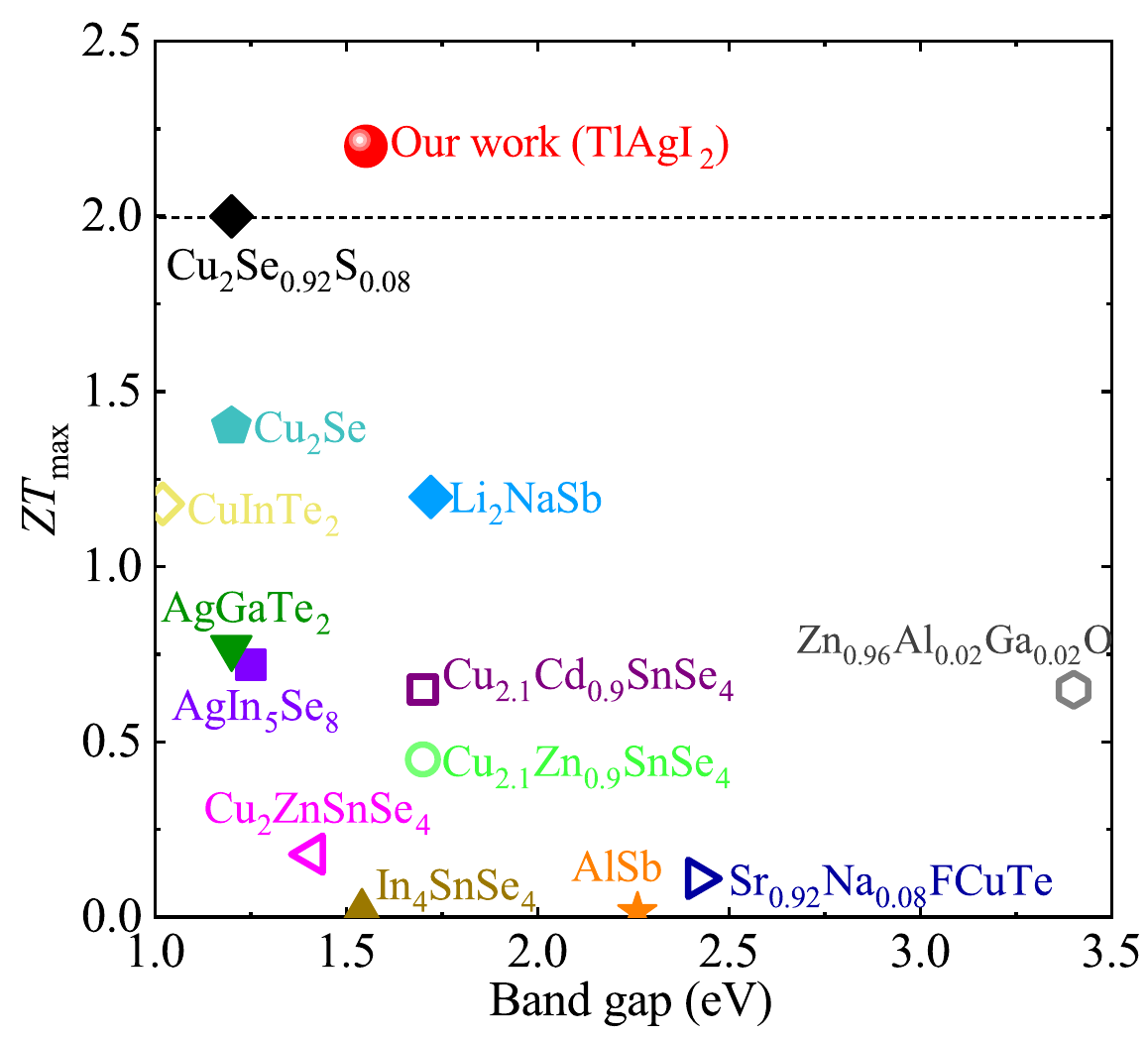}
\caption{
Some bandgap and thermoelectric data from experiments and calculations at reaching $ZT_{max}
$ temperatures. In$_4$SnSe$_4$~\cite{shi2022promising}, Li$_2$NaSb~\cite{xing2017electronic}, AgIn$_5$Se$_8$~\cite{cui2012microstructure}, Cu$_2$ZnSnSe$_4$~\cite{liu2009improved}, CuInTe$_2$~\cite{liu2012ternary}, Cu$_{2.1}$Cd$_{0.9}$SnSe$_4$ and Cu$_{2.1}$Zn$_{0.9}$SnSe$_4$~\cite{shi2009thermoelectric}, Sr$_{0.92}$Na$_{0.08}$FCuTe and Sr$_{0.94}$Na$_{0.06}$FCuTe ~\cite{JIANG2021122169}, AlSb~\cite{shawon2019mechanical}, Cu$_2$Se and Cu$_2$Se$_{0.92}$S$_{0.08}$~\cite{lu2015multiformity}, AgGaTe$_2$~\cite{su2019high}, Zn$_{0.96}$Al$_{0.02}$Ga$_{0.02}$~\cite{pilasuta2013characterization}. 
\label{fig7}}
\end{figure}

Based on our investigation of existing wide bandgap thermoelectric materials, including pure semiconductors and doping-modified multifunctional semiconductor materials illustrated in Fig.~\ref{fig7}, it is noteworthy that most of the thermoelectric materials with a bandgap exceeding 1.0 eV exhibit $ZT$ values below 1.0. A notable exception is Cu$_2$Se, where doping with sulfur element elevates the ZT from 1.4 to 2.0. Consequently, our findings establish that the $ZT$ value of 2.20 for TlAgI$_2$ takes a leading position in the bandgap range of 1.0 eV to 3.5 eV. This suggests that TlAgI$_2$ holds significant promise as a potential thermoelectric material at high temperatures. Furthermore, there is potential to enhance the $ZT$ value through doping. Therefore, it is advisable to explore the application of wide bandgap semiconductors in high-temperature thermoelectric materials.



\section{IV. CONCLUSIONS}



In summary, we employed first-principles calculations, the self-consistent phonon (SCPH) theory, and Boltzmann transport equations to investigate the thermal and electrical transport properties of TlAgI$_2$. The results revealed distinctive effects of quartic anharmonicity and coherent phonons on lattice thermal conductivity. Key findings include:

(i) The study highlighted the significant contributions of four-phonon processes, anharmonicity phonon renormalization, and coherent phonons in achieving ultralow thermal conductivity. These factors are crucial in theoretical predictions for high thermoelectric performance materials.
(ii) The observed low thermal conductivity in TlAgI$_2$ is attributed to antibonding states of Ag 4d and I 5p orbitals below the Fermi level, along with the bonding hierarchy of ionic, covalent, and coordination interactions. Strong Ag-I coordination interactions lead to a large valence band state density. Favorable electrical transport properties are linked to high energy band asymmetry and a wide bandgap, countering bipolar effects, especially at high temperatures.
(iii) TlAgI$_2$ emerges as a potential candidate for thermoelectric applications due to its ultralow thermal conductivity and favorable electrical transport properties.
This research contributes to advancing our understanding of the thermal and electrical properties of TlAgI$_2$, offering guidance for the exploration of materials with wide bandgaps for high-temperature thermoelectric applications.

\section{ACKNOWLEDGMENTS}
This work is sponsored by the Key Research and Development Program of the Ministry of Science and Technology (No.2023YFB4604100).
We acknowledge the support from the National Natural Science Foundation of China 
(No.12104356 and No.52250191, No.22103099), 
the Opening Project of Shanghai Key Laboratory of Special Artificial Microstructure Materials 
and Technology (No.Ammt2022B-1), and the Fundamental Research Funds for the Central 
Universities. 
W. Shi acknowledges the support from the Guangzhou Science and Technology Plan Project (202201011155).
We also acknowledge the support by HPC Platform, Xi’an Jiaotong University. 

 \bibliography{References} 
 \end{document}